\documentclass[a4paper,11pt]{article}
\usepackage[dvips]{graphicx}
\usepackage{amsmath}
\usepackage{amsfonts}
\usepackage{amssymb}

\date{}

\begin{document}

\title{Fluorescence enhancement factors on optical antennas: enlarging the experimental values
without changing the antenna design}
\author{J\'{e}r\^{o}me Wenger \\Institut Fresnel, CNRS, Aix Marseille Universit\'{e}, Ecole
Centrale Marseille,\\ Campus St J\'{e}r\^{o}me, 13013 Marseille,
France\\ email: jerome.wenger@fresnel.fr}

\maketitle

\begin{abstract}
Plasmonic antennas offer promising opportunities to control the
emission of quantum objects. As a consequence, the fluorescence
enhancement factor is widely used as a figure of merit for a
practical antenna realization. However, the fluorescence
enhancement factor is not an intrinsic property of the antenna. It
critically depends on several parameters, some of which are often
disregarded. In this contribution, I explore the influence of the
setup collection efficiency, emitter's quantum yield and
excitation intensity. Improperly setting these parameters may
significantly alter the enhancement values, leading to potential
misinterpretations. The discussion is illustrated by an antenna
example of a nanoaperture surrounded by plasmonic corrugations.
\end{abstract}

\section{Introduction}

Plasmonic antennas are receiving a large interest to interface
light with nanoscale quantum emitters on dimensions much beyond
the optical wavelength \cite{NovotnyRev11,NovotnyRev09}. Recent
developments involve squeezing light into nanoscale volumes
\cite{BrongersmaRev}, enhancing the excitation and emission rate
of individual emitters
\cite{Anger06,Kuhn06,Muskens07,Moerner09,LakowiczRev}, tuning the
luminescence spectrum \cite{Feldmann08,Gomez09}, polarization
\cite{Moerland08} and directivity properties
\cite{Gersen00,Taminiau08,Kuhn08,Curto10,Kall11}. Several
plasmonic systems are being investigated to enhance the
luminescence emission of fluorescent molecules or quantum dots,
such as metallic nanoparticles
\cite{Anger06,Kuhn06,lakowicz07,hohenester07,geddes07,Colas08},
core-shell particles \cite{tam07}, thin films
\cite{1Barnes,enderlein05}, nanoantennas
\cite{Muskens07,Moerner09,Curto10,Rogobete07}, nanowires
\cite{Kall11}, nanoporous gold \cite{biteen05}, nanopockets
\cite{liu06}, metallic gratings \cite{hung06}, nanoaperture arrays
\cite{blair03}, and single nanoapertures \cite{Davy08,Wenger08}. A
general review on surface-enhanced fluorescence can be found in
reference \cite{Fort08}.

A natural question while performing experiments on
nanoantenna-enhanced luminescence deals with the quantification of
the luminescence enhancement factor $\eta_F$, which is commonly
defined as the ratio of the detected radiation power per emitter
with the antenna to the reference radiation power per emitter
without the antenna. $\eta_F$ determines how many extra photons
are detected for each emitter thanks to the use of the optical
antenna. It is well known that this factor critically depends on
several parameters: the antenna material and geometry, its
spectral resonance and overlap with the emitter's absorption and
luminescence spectra, as well as the emitters's orientation and
location respective to the antenna \cite{NovotnyBook}. These many
parameters often hide the influence of other parameters: the
collection efficiency used in the experiments, the emitter's
quantum yield in the absence of the antenna, and the excitation
intensity respective to the saturation process. These last three
parameters are more technically oriented, and depend on the
specific experimental implementation. Therefore, they have
received less attention from theoretical investigations. However,
as I will show below, these parameters have a major influence on
the measured values of the luminescence enhancement factor.
Improperly setting these parameters may significantly alter the
value found for the luminescence enhancement factor, leading to
potential experimental pitfalls.

\begin{figure}[t]
\begin{center}
\includegraphics[width=7cm]{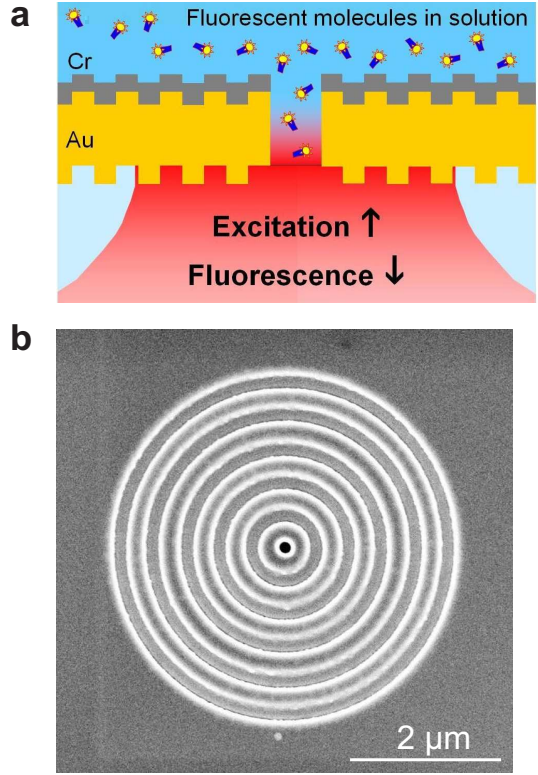}
\caption{Corrugated aperture antenna used here to illustrate the
influence of experimental parameters on the measured fluorescence
enhancement factor. (a) Experimental configuration to probe
molecules randomly diffusing in aqueous solution. (b)  Scanning
electron microscope image of the antenna with five corrugations.
Adapted with permission from \cite{AouaniNL}. Copyright 2011
American Chemical Society.}\label{Fig:intro}
\end{center}
\end{figure}

In this contribution, I explore the influence of the collection
efficiency, molecular quantum yield and excitation intensity on
the fluorescence enhancement factor. Several rules are derived to
maximize the enhancement factor by tuning the experimental
conditions without affecting the antenna's design. Using the
fluorescence enhancement factor as a figure of merit to compare
between different antenna designs has to be done with caution.
Analytical formulas are illustrated by a practical antenna example
made of a single nanoaperture surrounded by five shallow grooves
in a gold film (Figure~\ref{Fig:intro}). This optical antenna can
significantly enhance the fluorescence count rates per molecule
and control the emission directivity, as demonstrated recently
\cite{AouaniNL}. Very much in the spirit of the review in
\cite{BarnesJOPA} about the comparison between experiments and
numerical simulations, I hope that this article may initiate some
reflection, and avoid any misleading interpretation.

\section{Designing the experiment to maximize the fluorescence enhancement factor}

A loophole in any measurement of the luminescence enhancement
factor deals with the normalization of the detected signal
\textit{per emitter}. This can be done by performing experiments
on single molecules directly \cite{Anger06,Kuhn06,Moerner09}, by
surface normalization \cite{biteen05,blair03}, or by calibrating
the number of emitters with fluorescence correlation spectroscopy
\cite{Davy08,Wenger08}. Hereafter, I consider that this
normalization issue is resolved, and that a reliable value for the
detected radiated power per emitter can be obtained. I also assume
that the signal per emitter has been averaged over several
positions and dipole orientations. The aim of this article is to
focus on the influence of experimental parameters on the measured
values for the fluorescence enhancement factor $\eta_F$.

\subsection{Theoretical background: fluorescence count rate per
molecule}

Throughout this article, the quantum emitter is modelled by a two
energy levels system. $k_{r}$ and $k_{nr}$ are the rate constants
for radiative and non-radiative transitions from the excited
singlet state to the ground state. The total deexcitation rate
from the excited singlet state is noted as $k_{tot} = k_r+ k_{nr}
$, which is the inverse of the excited state lifetime $\tau$.
$\phi = k_r/k_{tot}= k_{r}/(k_{r}+k_{nr})$ is the quantum yield.
$\sigma I_e$ is the excitation rate, where $\sigma$ denotes the
excitation cross-section and $I_e$ the excitation intensity. Under
steady-state conditions, the fluorescence count rate per molecule
$CRM$ is given by \cite{Fluobouquin}
\begin{equation}\label{Eq:FM}
  CRM =  \kappa \, \phi \, \frac{\sigma I_e}{1+I_e/I_s}.
\end{equation}
We note $\kappa$ the light collection efficiency, and $I_s =
k_{tot}/\sigma$ the saturation intensity.

Equation~\ref{Eq:FM} takes two limits for the extreme regimes of
weak excitation ($I_e \ll I_s$) and fluorescence saturation ($I_e
\gg I_s$). In the weak excitation regime, the $CRM$ reduces to
\begin{equation}\label{Eq:FMlow}
  CRM =  \kappa \, \phi \, \sigma \, I_e \,\,\,\,\,\,\,\,\,\,\,\,\, (I_e \ll
  I_s)
\end{equation}
which indicates that the fluorescence rate per molecule is
proportional to the collection efficiency, the quantum yield, and
the excitation intensity. This expression appears to be the one
commonly used in fluorescence spectroscopy and microscopy applied
to the life sciences \cite{LakowiczBook}.

In the saturation regime $I_e \gg I_s$, Eq.~(\ref{Eq:FM}) reduces
to
\begin{equation}\label{Eq:FMsat}
  CRM = \kappa \, \phi \, \sigma \, I_s = \kappa \, k_r \,\,\,\,\,\,\,\,\,\,\,\,\, (I_e \gg I_s)
\end{equation}
which indicates that the fluorescence rate per molecule at
saturation is determined only by the radiative rate and the
collection efficiency, and is of course independent on the
excitation rate. This expression is generally used for
single-photon sources used for quantum communication purposes,
such as quantum cryptography \cite{Alexios1,Alexios2}.

\subsection{Excitation intensity}

From Eq.~(\ref{Eq:FM}), it is apparent that the detected
fluorescence rate per emitter -- hence the fluorescence
enhancement factor -- bears a complex dependence upon the
excitation intensity $I_e$. In the weak excitation regime (below
the transition to saturation), the $CRM$ is linearly proportional
with $I_e$, and the fluorescence enhancement factor $\eta_F$ can
therefore be expressed as
\begin{equation}\label{Eq:etaFlow}
  \eta_F = \frac{CRM^{\ast}}{CRM} = \frac{\kappa^{\ast}}{\kappa} \, \frac{\phi^{\ast}}{\phi} \frac{I_e^{\ast}}{I_{e}} \, \,\,\,\,\,\,\,\,\,\,\,\,\, (I_e \ll
  I_s).
\end{equation}
The superscript $^{\ast}$ denotes the presence of the antenna. In
the weak excitation regime, the fluorescence enhancement factor is
the product of the enhancements in the collection efficiency, the
quantum yield and the excitation intensity (here we assume that
the nanoantenna does not modify significantly the fluorophore's
absorption cross section: $\sigma^{\ast}=\sigma$).

Equation~(\ref{Eq:etaFlow}) can be rewritten in a slightly
different manner to introduce the gains in the radiative rate
$k_r^{\ast}/k_r$ and the total fluorescence lifetime reduction
$k_{tot}^{\ast}/k_{tot}$:
\begin{equation}\label{Eq:etaFlow2}
  \eta_F = \frac{\kappa^{\ast}}{\kappa} \, \frac{k_{r}^{\ast}}{k_{r}} \, \frac{k_{tot}}{k_{tot}^{\ast}} \, \frac{I_e^{\ast}}{I_{e}} = \frac{\kappa^{\ast}}{\kappa} \, \frac{k_{r}^{\ast}}{k_{r}} \, \frac{\tau^{\ast}}{\tau} \, \frac{I_e^{\ast}}{I_{e}} \, \,\,\,\,\,\,\,\,\,\,\,\,\, (I_e \ll
  I_s).
\end{equation}
This equation shows that the fluorescence enhancement factor is
proportional to the gains in the radiative rates $k_r$, and
\textit{inversely} proportional to the total deexcitation rates
$k_{tot}$. Since $k_{tot}^{\ast}/k_{tot}=\tau/\tau^{\ast}$ is also
the reduction in the fluorescence lifetimes, a strong reduction in
the fluorescence lifetimes (sometimes also referred to as Purcell
factor) is actually detrimental to the fluorescence enhancement
factor. Clearly, the fluorescence lifetimes reduction (Purcell
factor) must \textit{not} be confused with the fluorescence count
rate enhancement (see also the discussion in \cite{KoenderinkOL}).
Observing a strong lifetime reduction due to the nanoantenna can
be related to an increase in the non-radiative transition rate
$k_{nr}^{\ast}$, mostly related to ohmic losses. This is the
so-called quenching effect, which is certainly not related to any
increase in the fluorescence count rate per emitter.

In the saturation regime, it is evident from Eq.~(\ref{Eq:FMsat})
that the fluorescence enhancement factor at saturation depends
only on the gains in collection efficiency and radiative rate:
\begin{equation}\label{Eq:etaFsat}
  \eta_{F} = \frac{\kappa^{\ast}}{\kappa} \,\frac{k_r^{\ast}}{k_r} \,\,\,\,\,\,\,\,\,\,\,\,\,\, (I_e \gg
  I_s)
\end{equation}
Interestingly, at saturation of the fluorescence process, the
fluorescence enhancement is found independent on the non-radiative
transition rate $k_{nr}$, which means that metal quenching might
not be an issue in this particular configuration. At saturation,
the sole figure of merit is the ability of the
emitter--nanoantenna system to radiate photons into the photonic
modes used for detection, which is quantified by $\kappa k_r$
(this rate is sometimes written $k_{em}= \kappa k_r$
\cite{Wenger08,AouaniNL}). Comparing the enhancement $\eta_F$ in
the saturation regime to the case of weak excitation, it turns out
that $\eta_F$ in the weak excitation regime is larger by a factor
$(I_{e}^{\ast}/I_{e})(\tau^{\ast}/\tau)$, which amounts to the
gain in the excitation intensity divided by the fluorescence
lifetime reduction. Depending on the interplay between excitation
enhancement and quenching losses, high fluorescence enhancement
values can be preferentially reached by working in the weak
excitation regime (most common cases), or at fluorescence
saturation (in the case of strong quenching losses). However, it
must be kept in mind that the saturation intensity increases when
the reduction of the fluorescence lifetime. Therefore reaching the
saturation regime will require more excitation power in the case
of strong quenching, that may even induce photodamage to the
molecule or to the structure.

\begin{figure*}[t]
\begin{center}
\includegraphics[width=17cm]{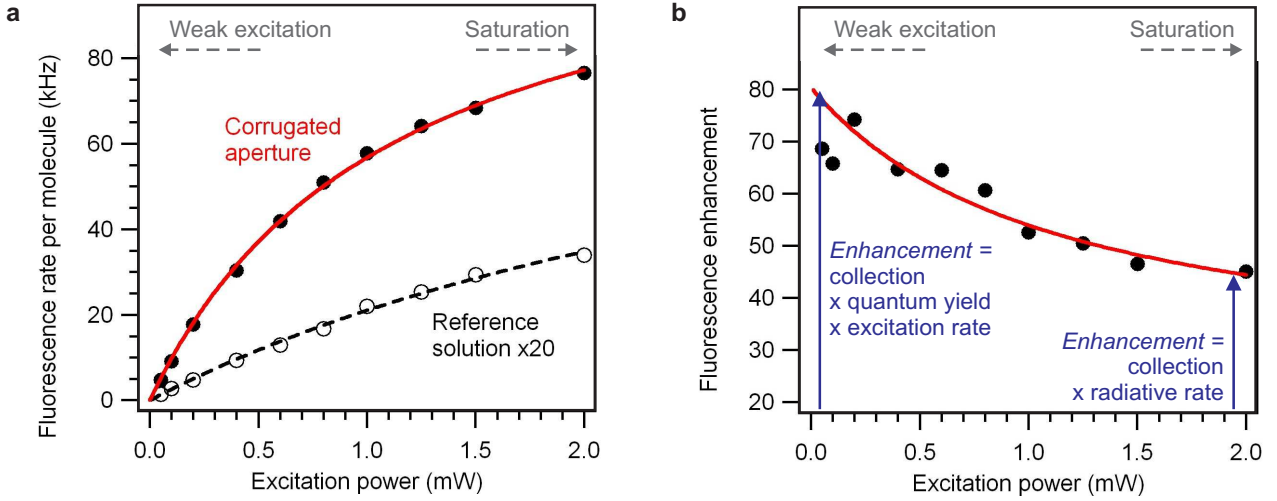}
\caption{(a) Average fluorescence count rate per molecule detected
in the case of the corrugated aperture presented in
Fig.~\ref{Fig:intro} (filled circles) and the reference solution
(empty circles). The molecules have a quantum yield in solution of
30\% (Alexa Fluor 647), and the numerical aperture used for both
excitation and collection is 0.5. (b) Fluorescence enhancement
factor for increasing excitation power. The arrows indicate the
two extreme cases of weak excitation ($I_e \ll I_{s}$) and
saturation ($I_e \gg I_{s}$). Figure adapted with permission from
\cite{AouaniNL}.}\label{Fig:Iex}
\end{center}
\end{figure*}

Figure~\ref{Fig:Iex} illustrates the dependence of the
fluorescence enhancement factor on the excitation intensity, in
the example case of the corrugated aperture displayed in
Fig.~\ref{Fig:intro}. The fluorescence count rates per molecule in
the case of the antenna and the reference solution are presented
in Fig.~\ref{Fig:Iex}(a) versus the excitation power, together
with numerical fits according to the general expression
Eq.~(\ref{Eq:FM}) (the $CRM$ for the reference solution has been
multiplied by an arbitrary 20x factor to ease viewing). A
transition between the regimes of weak excitation and saturation
are clearly seen for excitation powers around 1~mW (excitation
made through a 0.5~NA objective, so the power to reach saturation
are significantly higher as compared to focusing with a high NA
objective). The transition towards saturation can be quantified by
the saturation intensity $I_s$. For the experiments in
Fig.~\ref{Fig:Iex}(a), we found $I_s^{\ast}=1.2$~mW for the
antenna and $I_s=3.4$~mW for the reference solution
\cite{AouaniNL} (in the case of a 1.2~NA objective, the saturation
intensities would be $I_s^{\ast}=130$~$\mu$W for the antenna and
$I_s=510$~$\mu$W for the reference solution).

Changing the excitation intensity has a strong influence on the
fluorescence enhancement factor, as seen in Fig.~\ref{Fig:Iex}(b).
In the weak excitation regime, the enhancement is maximum ($\sim
80\times$ for this example), while it decreases to $\sim 40\times$
when saturation is reached. This can be directly explained by the
different contributions of the gains in radiative rate, quantum
yield, and excitation intensity. $\eta_F$ in the weak excitation
regime is larger by a factor
$(I_{e}^{\ast}/I_{e})(\tau^{\ast}/\tau) \simeq 2$, which
corresponds well to the gain in the excitation intensity ($\approx
5.5$) divided by the fluorescence lifetime reduction ($\approx 2$)
as measured separately \cite{AouaniNL}.

\subsection{Emitter's reference quantum yield without the antenna}

In this section, we take a closer look at the influence of the
emitter's quantum yield $\phi = k_{r}/(k_{r}+k_{nr})$ on the
fluorescence enhancement factor. The basic question is how shall
we choose $\phi$ for the reference solution (without the antenna)
so as to maximize the fluorescence enhancement? Hereafter, the
excitation intensity is set to the weak excitation regime $I_e \ll
I_s$ (in the saturation regime, the fluorescence enhancement does
not depend directly on $\phi$, so the discussion is useless).

With the nanoantenna, the quantum yield is modified to
$\phi^{\ast} =k_{r}^{\ast}/(k_{r}^{\ast}+ k_{nr}+k_{abs}^{\ast})$,
where a new non-radiative decay route $k_{abs}^{\ast}$ is
introduced to take into account the ohmic losses into the metal
and non-radiative energy transfers to the free electrons in the
metal. It is also assumed that the non-radiative rate $k_{nr}$ is
not affected by the antenna. After some basic algebra,
Eq.~(\ref{Eq:etaFlow}) can be rewritten:
\begin{equation}\label{EqEtaPhi}
  \eta_F = \frac{\kappa^{\ast}}{\kappa} \,
  \frac{k_{r}^{\ast}}{k_{r}} \, \frac{I_e^{\ast}}{I_{e}} \, \frac{1}{(1-\phi) + \phi \zeta}
\end{equation}
with $\zeta = (k_r^{\ast} + k_{abs}^{\ast})/k_{r}$. In the limit
of a ``poor'' emitter $\phi \ll 1$, and $\phi \zeta \ll 1$,
Eq.~(\ref{EqEtaPhi}) resumes to
\begin{equation}\label{EqEtaPhiLow}
  \eta_F = \frac{\kappa^{\ast}}{\kappa} \, \frac{k_{r}^{\ast}}{k_{r}} \,\frac{I_e^{\ast}}{I_{e}} \,\,\,\,\,\,\,\,\,\,\,\,\,\,
  (\phi \ll 1)
\end{equation}
which is the product of the gains in collection efficiency,
radiative rate and excitation intensity (this value can also be
seen as the enhancement factor found at saturation times the gain
in excitation intensity).

In the case of a perfect emitter $\phi \simeq 1$,
Eq.~(\ref{EqEtaPhi}) resumes to
\begin{equation}\label{EqEtaPhiHigh}
  \eta_F = \frac{\kappa^{\ast}}{\kappa} \,\frac{I_e^{\ast}}{I_{e}} \,\,\,\,\,\,\,\,\,\,\,\,\,\,
  (\phi \simeq 1)
\end{equation}
if we assume that $k_r^{\ast} \gg k_{abs}^{\ast}$, meaning that
the antenna has a large efficiency. Thus for a perfect emitter,
the fluorescence enhancement at weak excitation intensity is given
by the product of the gains in collection efficiency and
excitation intensity. The fluorescence enhancement in the case of
a poor emitter is larger by a factor $k_r^{\ast}/k_r$, indicating
that in order to maximize the value for $\eta_F$, one should
preferentially select emitters with rather low quantum yields (as
long as the experimental signal-to-noise ratio is sufficiently
high to provide for reliable measurements).

\begin{figure}[t]
\begin{center}
\includegraphics[width=8.3cm]{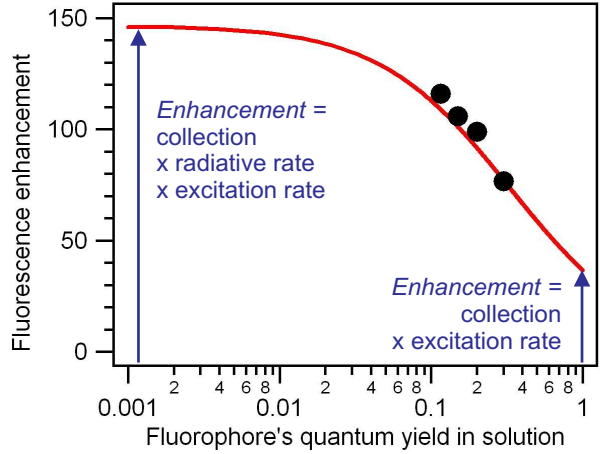}
\caption{Fluorescence enhancement factor versus the dye's quantum
yield $\phi$ in solution (taken without the antenna), in the case
of weak excitation. The arrows indicate the two extreme cases of
perfect dye ($\phi=1$) or poor emitter ($\phi\ll1$). The
experimental data is taken from \cite{AouaniNL}.}\label{Fig:phi}
\end{center}
\end{figure}

Figure~\ref{Fig:phi} illustrates this discussion, based on the
modification of the fluorescence properties calibrated in
\cite{AouaniNL}. $\eta_F$ grows as the reference quantum yield
$\phi$ is decreased, up to a plateau for $\phi < 0.02$. Again, the
maximum value for the fluorescence enhancement in this example can
be significantly different if a perfect dye is used $\eta_F
\approx 35$, or if a poor emitter is chosen $\eta_F \approx 150$.

\subsection{Luminescence collection efficiency}

All the formulas presented here for the fluorescence enhancement
factor are proportional to the gain in collection efficiency
$\kappa^{\ast}/\kappa$, independently on the choice for excitation
intensity or quantum yield. This is a direct consequence that
obviously only the photons that are collected by the setup
contribute to the detected fluorescence. Therefore, to maximize
the fluorescence enhancement factor, one has to maximize the
collection efficiency gain $\kappa^{\ast}/\kappa$. This can be
done by tuning the antenna so as to maximize the directivity
\cite{Curto10,TaminiauNJP08,AouaniNL2}, \textit{and} by minimizing
the collection efficiency $\kappa$ used for the reference. In
other words, to maximize $\eta_F$, an objective with low numerical
aperture should be used in accordance to the peak angular emission
of the antenna's radiation pattern.

\begin{figure}[t]
\begin{center}
\includegraphics[width=8.3cm]{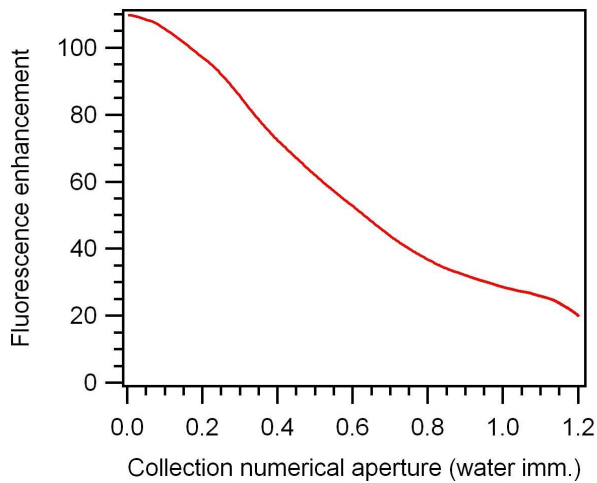}
\caption{Influence of the numerical aperture used for collection
on the fluorescence enhancement factor. Computation derived from
the experimental data published in \cite{AouaniNL}, in the case of
weak excitation.}\label{Fig:NA}
\end{center}
\end{figure}

Figure~\ref{Fig:NA} displays the influence of the numerical
aperture used for fluorescence collection on the detected
fluorescence enhancement factor. The antenna used here is again
the corrugated aperture presented in Fig.~\ref{Fig:intro}. It was
found that the fluorescence radiation pattern with the antenna
presented a peak in the direction normal to the sample plane with
an angular divergence of $\pm15^{\circ}$. Figure~\ref{Fig:NA}
presents the computed fluorescence enhancement factor integrated
over the whole collection NA. Minimizing the collection NA to
values below 0.1 maximizes the fluorescence enhancement factor up
to values $\sim 110$, whereas a large collection NA of 1.2 reduces
the effective enhancement factor to $\sim 20$ as an effect of
angular averaging.

\section{Discussion}

From the aforementioned investigations on the role of the
different experimental parameters on the measured fluorescence
enhancement factor, a general ``rule of thumb'' can be derived:
the lower the reference without the antenna, the higher the
enhancement factor. In other words, to obtain high enhancement
factors, one should preferentially select a weak excitation
regime, a dye with low quantum yield and a low collection NA. Of
course, this comes at the expense of lower signal (at least for
the reference case) and higher noise.

\begin{figure*}[t]
\begin{center}
\includegraphics[width=17cm]{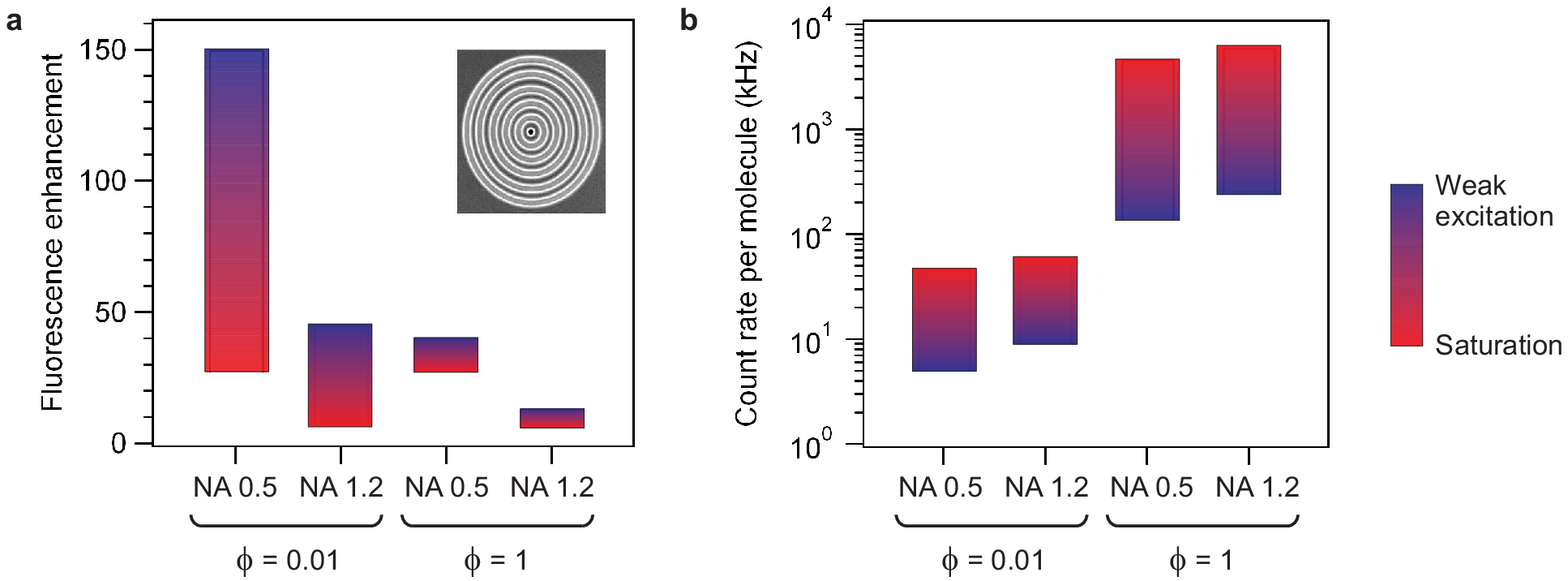}
\caption{Fluorescence enhancement factor (a) and fluorescence
detection rate per emitter (b) for different numerical apertures
(NA) used for collection, different quantum yield $\phi$ of the
emitter in solution (in the absence of the antenna) and in the
limit of weak excitation (blue) or fluorescence saturation (red).
The antenna used here is the corrugated aperture with five grooves
described in Fig.~\ref{Fig:intro}, and calibrated in
\cite{AouaniNL}.}\label{Fig:bilan}
\end{center}
\end{figure*}

To illustrate the broad range of fluorescence enhancement factors
that can be measured \textit{using the same nanoantenna design},
different computations have been performed using the antenna
presented in Fig.~\ref{Fig:intro} with either 0.5 or 1.2~NA
microscope objective, 0.01 or 1 quantum efficiencies and in the
limits of weak excitation or fluorescence saturation. The results
are graphically presented in Fig.~\ref{Fig:bilan}(a). In the
extreme case of low reference without the antenna (0.5~NA,
$\phi=0.01$, weak excitation), the fluorescence enhancement is
maximum $\sim 150$, while it is minimum $\sim 6$ when the
reference is the highest (1.2~NA, $\phi=1$, saturation).

This clearly shows that the fluorescence enhancement factor is
\textit{not} an absolute figure of merit for a given nanoantenna.
This is in strong opposition with the directivity, which is
another figure of merit commonly used to characterize the quality
of an antenna design, and which can be intrinsic to the antenna
(if properly measured \cite{NovotnyRev09,AouaniNL,AouaniNL2}).

The fluorescence enhancement factor is an intuitive metric that is
commonly used in surface-enhanced (or metal-enhanced)
fluorescence. The experimental setup can be adequately tuned so as
to maximize the measured values of $\eta_F$. However, it should be
kept in mind that the final goal in the detection of molecules or
the generation of single photons is to realize bright sources out
of single quantum emitters \cite{Sandoghdar11}. Thus the true
figure of merit is not the enhancement factor (``how much do we
\textit{gain}'') but instead the fluorescence count rate per
emitter (``how much do we \textit{have}''). High enhancement
factors should not be confused with bright sources. To illustrate
this, Figure~\ref{Fig:bilan}(b) displays the detected count rate
per molecule corresponding to the cases in (a). To maximize the
detection rate, one has to select the conditions of high
collection NA, high quantum yield, and high excitation intensity
(saturation). These conditions correspond to the ones leading to
the minimum fluorescence enhancement factor in
Fig.~\ref{Fig:bilan}(a). Reciprocally, the conditions found for
maximum enhancement lead to the minimum detection rates. Taking a
low emission for reference and enhancing it (a lot) does not
necessarily compensate for the low detection rate to start with.
Of course, all this discussion strongly depends on the final
application of the nanoantenna--emitter system, and the initial
photophysical properties of the emitter.

\section{Conclusion}

This article has explored the influence of several parameters on
the fluorescence enhancement factor with an optical antenna:
excitation intensity, emitter's quantum yield, and collection
efficiency. General rules have been derived to obtain high
enhancement factors by tuning the experimental conditions without
affecting the antenna's design. Experimental conditions leading to
a low reference signal without the antenna should be
preferentially selected to maximize $\eta_F$. This corresponds to
weak excitation regime, dye with low quantum yield and low
collection NA. General remarks can be drawn from the discussion:
(i) the fluorescence enhancement factor is not an absolute figure
of merit for a given optical antenna design, (ii) the fluorescence
lifetime reduction (Purcell factor) must not be confused with the
fluorescence count rate enhancement, and (iii) high enhancement
factors do not necessarily indicate bright photon sources.

\section*{Acknowledgements}

I am deeply indebted to many at the Fresnel Institute and the
Laboratoire des Nanostructures at the Institut de Science et
d'Ing\'{e}nierie Supramol\'{e}culaires. I would like to thank
Herv\'{e} Rigneault, Pierre-Fran\c{c}ois Lenne and Thomas Ebbesen
for stimulating the early stages of this work, Oussama Mahboub and
Elo\"{\i}se Devaux for fabricating the sample presented in
Figure~\ref{Fig:intro}, and Heykel Aouani for performing the
measurements. I also would like to acknowledge stimulating
discussions with Steve Blair, Davy G\'{e}rard, Nicolas Bonod,
Brian Stout and Evgeny Popov. To Nicolas, I owe a special thank
for stimulating this work consecutively to the Second Summer
School on Plasmonics held on Porquerolles in Fall 2011. Lastly, I
would like to thank Didier Wampas for inspiring part of this work.

The research leading to these results has received funding from
the European Research Council under the European Union's Seventh
Framework Programme ERC Starting Grant agreement 278242. The
research has been conducted in the scope of the CNRS-Weizmann NaBi
European associated laboratory.


\end{document}